 \documentclass[12pt, draftclsnofoot, onecolumn]{IEEEtran} 
\usepackage[LGR,T1]{fontenc}
\usepackage[latin9]{inputenc}
\usepackage{enumitem}
\usepackage{amsmath, amsfonts, amssymb, mathtools}
\usepackage{amssymb}
\usepackage{graphicx}
\usepackage{wasysym}

\usepackage{textgreek}
\usepackage{textalpha}

\makeatletter

\DeclareRobustCommand{\greektext}{%
  \fontencoding{LGR}\selectfont\def\encodingdefault{LGR}}
\DeclareRobustCommand{\textgreek}[1]{\leavevmode{\greektext #1}}
\ProvideTextCommand{\~}{LGR}[1]{\char126#1}

\usepackage{graphicx}
\usepackage{graphicx, caption, subcaption}
\usepackage[ruled,vlined]{algorithm2e} % Include common option

\usepackage{algpseudocode}

\usepackage{relsize}
\usepackage{ tabularx}

\usepackage{array}     % For custom column formatting
 \usepackage{booktabs}  % For professional-looking tables

\usepackage{newtxtext,newtxmath} % Times-like fonts
\usepackage[english]{babel}

\usepackage[numbers,square,comma,sort&compress]{natbib}

\usepackage{adjustbox, listings, import, varioref}

\makeatletter
\long\def\@makecaption#1#2{
   \ifx\@captype\@IEEEtablestring
       \footnotesize\begin{center}{\normalfont\footnotesize #1}\\
       {\normalfont\footnotesize\scshape #2}\end{center}%
       \@IEEEtablecaptionsepspace
   \else
       \@IEEEfigurecaptionsepspace
       \setbox\@tempboxa\hbox{\normalfont\footnotesize {#1.}~~ #2}%
       \ifdim \wd\@tempboxa >\hsize
           \setbox\@tempboxa\hbox{\normalfont\footnotesize {#1.}~~ }%
           \parbox[t]{\hsize}{\normalfont\footnotesize \noindent\unhbox\@tempboxa#2}%
       \else
           \hbox to\hsize{\normalfont\footnotesize\hfil\box\@tempboxa\hfil}
       \fi
   \fi
}
\makeatother

\makeatother

\begin{document}

\title{Adaptive Interference Management for Enhancing RIS-Assisted NOMA Systems through Satellite-Terrestrial Links}

\author{
Muhammad Khalil,~\IEEEmembership{Member,~IEEE}, 
Ke Wang,~\IEEEmembership{Senior Member,~IEEE},\\ 
and Jinho Choi,~\IEEEmembership{Fellow,~IEEE}%
\thanks{
M. Khalil and K. Wang are with the School of Engineering, RMIT University, Melbourne, VIC 3000, Australia (e-mail: muhammad.khalil@rmit.edu.au; ke.wang@rmit.edu.au).\\
J. Choi is with the School of Electrical and Mechanical Engineering, University of Adelaide, Adelaide, SA 5005, Australia (e-mail: jinho.choi@adelaide.edu.au).}
}

\maketitle

\begin{abstract}
This paper introduces an elevation-aware adaptive interference management framework for Reconfigurable Intelligent Surface (RIS)-assisted downlink Non-Orthogonal Multiple Access (NOMA) systems in integrated low Earth orbit (LEO) satellite\textendash terrestrial networks. In these systems, satellite interference varies significantly with the elevation angle due to changes in path loss and shadowing, presenting a major challenge for reliable communication. To address this, we propose an analytically modeled elevation-dependent weighting factor, $\alpha(\theta)$, expressed as a sigmoid function and dynamically tuned in real time by a Proportional-Integral (PI) controller. This factor effectively captures both large-scale and small-scale channel impairments, allowing the system to adapt to fluctuating interference conditions. Leveraging $\alpha(\theta)$, we design an adaptive successive interference cancellation (SIC) strategy that automatically prioritizes the cancellation of satellite or terrestrial interference based on the current elevation angle, eliminating unnecessary decoding stages when satellite interference is weak. Additionally, we develop a three-mode RIS codebook, including terrestrial-priority, satellite-priority, and balanced modes, which is selected according to the value of $\alpha(\theta)$ and further optimized online using lightweight gradient-based methods to ensure optimal phase adaptation.

Extensive simulations with realistic operating parameters show that the proposed framework achieves up to 20\% higher system throughput compared to traditional static RIS schemes. These findings demonstrate the practical benefits of combining elevation-aware RIS control with adaptive interference management, offering a robust solution for enhancing connectivity and spectral efficiency in next-generation satellite\textendash terrestrial NOMA networks.
\end{abstract}
\section{Introduction}
Low Earth Orbit (LEO) satellites have been widely recognized as fundamental enablers of sixth-generation infrastructure, offering distinct advantages over traditional Geostationary Earth Orbit (GEO) and Medium Earth Orbit (MEO) systems. Operating at altitudes of only a few hundred kilometers, LEO satellites substantially reduce communication latency compared to their higher-orbit counterparts \cite{You2020}. For example, a LEO satellite positioned at approximately 600 km can achieve a round-trip communication delay of less than 30 milliseconds, whereas a GEO satellite typically exhibits a delay exceeding 250 milliseconds \cite{Tekbiyik2022}. In addition to low latency, LEO systems support higher frequency reuse and facilitate the deployment of large-scale "mega-constellations," which provide flexible, scalable coverage and improved link budgets due to their smaller coverage footprints. These features collectively make LEO constellations ideally suited to deliver the low-latency, high-capacity, and globally accessible connectivity envisioned for 6G networks \cite{Darwish2022}.

Despite these advantages, LEO-to-ground communications face significant challenges stemming from the unique dynamics of LEO satellites. The high mobility of LEO satellites leads to large Doppler shifts and rapidly changing link geometry, complicating signal acquisition and tracking. Each satellite is only in view of the ground receiver for a short period, necessitating frequent handovers and robust link maintenance strategies. LEO satellites also have limited onboard power and hardware resources due to size and weight constraints, which restrict their transmission power and processing capabilities. Additionally, downlink signals can suffer severe attenuation, as the free-space path loss is significant over hundreds of kilometers; at millimeter-wave frequencies, atmospheric absorption and rain can further weaken the signal. In urban environments, blockage by buildings can disrupt the line-of-sight (LoS) path. Notably, the satellite's elevation angle $\theta$ (the angle of the satellite above the horizon) strongly affects link quality, as low elevation angles result in a longer and more obstructed path, leading to heavier shadowing and a lower probability of a clear LoS \cite{Zhu2022a,Khalil2025}. For example, at $\theta\:\approx10^{{^\circ}}$ in a dense city, the chance of an unobstructed LoS link can be as low as 28\%, whereas it is about 78\% in open rural areas \cite{Toka2024}. Such conditions lead to poor link budgets and signal outages if not mitigated. These challenges highlight the need for adaptive solutions that accommodate the dynamic nature of LEO satellite communication links \cite{Lv2024}.

One promising approach to improving LEO-to-ground communication is the integration of a terrestrial auxiliary system using Reconfigurable Intelligent Surfaces (RIS) \cite{Khalil2022}. An RIS is a planar surface composed of sub-wavelength elements that can adjust the phase (and sometimes the amplitude) of incident signals. By doing so, an RIS can dynamically shape electromagnetic wave propagation, such as   reflecting a satellite's signal toward a target receiver along a desired path \cite{DiRenzo2020}. In the LEO downlink context, a ground-deployed RIS (e.g., on a building facade or high terrain) can establish a favorable indirect path that complements the direct satellite-to-user link. This capability allows the network to overcome obstacles and focus signal power where needed \cite{Khalil2024}, which is crucial for preventing communication interruptions in urban areas where signals are typically obstructed by physical barriers \cite{Han2018}. Importantly, RISs achieve this with nearly passive hardware, introducing negligible additional noise and consuming minimal power, as they largely reflect signals rather than retransmitting them \cite{Basar2019,Khalil2024}. Therefore, the integration of RIS technology with LEO satellites offers a cost-effective method to address the aforementioned attenuation and blockage issues while requiring minimal additional ground infrastructure \cite{Averty2004}. 

In parallel, Non-Orthogonal Multiple Access (NOMA) has been proposed as an advanced multiple access scheme to improve spectral efficiency. Unlike orthogonal access, NOMA allows multiple users to share the same time-frequency resources by superimposing their signals at different power levels and employing successive interference cancellation (SIC) at receivers to decode the layered signals. This power-domain multiplexing
increases throughput, albeit at the cost of interference that must be managed through careful power allocation and decoding order design \cite{Choi2014}. Integrating RIS with NOMA provides a new synergy to optimize spectral efficiency and interference management. By jointly controlling the propagation environment (via RIS) and the power domain (via NOMA), the system gains additional degrees of freedom to enhance desired signals and suppress interference for each user \cite{Randrianantenaina2020}. In essence, RIS can improve signal quality on the fly (e.g. by beam focusing or redirecting signals), while NOMA maximizes spectral utilization by serving multiple users on the same resources \cite{Hou2020}. Optimized coordination of RIS beamforming with NOMA power allocation and SIC order enables the network to enhance each user's intended signal and reduce inter-user interference, achieving higher overall spectral efficiency than either technology could achieve independently \cite{Li2023a}.

While integrating NOMA, RIS, and LEO satellite links is often cited as a promising approach for achieving high-capacity, wide-area connectivity in next-generation wireless systems \cite{Khan2023}, the present work departs from this conventional integration model. In our framework, the LEO satellite does not serve as a data link; rather, it acts solely as an uncontrollable external interference source whose impact on terrestrial users varies sharply with the satellite \textendash user elevation angle $\theta$ \cite{Li2002}. As the satellite moves in its orbit, the elevation angle changes continuously, causing significant fluctuations in path loss, shadowing, and received interference power \cite{GongoraTorres2022}. At low $\theta$, the satellite signal experiences severe atmospheric attenuation, while at high $\theta$, it becomes a dominant line-of-sight interferer.

These dynamic elevation-dependent interference conditions create a unique challenge for terrestrial RIS-NOMA systems, as static RIS configurations and fixed SIC orders cannot adapt to rapid changes in satellite interference. Recent terrestrial NOMA studies demonstrate that dynamic SIC ordering based on instantaneous channel conditions can significantly improve performance \cite{Ding2020}, emphasizing the necessity for elevation-aware adaptation when LEO satellite signals serve as interference. This consideration is especially important given the rapid deployment of modern LEO constellations for extended coverage and regulatory compliance \cite{Lin2021}, making their signals an unavoidable aspect of future terrestrial network environments \cite{Juan2022}. 

To address this challenge, we introduce an elevation-dependent weighting factor $\alpha(\theta)\in\left[0,1\right]$, which explicitly quantifies the instantaneous satellite interference power relative to terrestrial signals. Physically, $\alpha(\theta)$ is near zero at low elevation angles, where satellite interference is minimal due to high path loss and shadowing, and approaches unity at high elevation angles, reflecting the dominance of line-of-sight satellite interference. Our primary contribution is the analytical modeling and real-time adaptation of this weighting factor. Specifically, $\alpha(\theta)$ is formulated as a smooth sigmoid function whose steepness and transition midpoint are dynamically adjusted by a proportional-integral (PI) controller. This controller operates on real-time measurements that capture both large-scale log-normal shadowing and small-scale fading, with the latter represented by a Gaussian mixture model. By estimating and dynamically updating $\alpha(\theta)$ at the terrestrial base station, our framework transforms complex elevation-dependent satellite interference dynamics into a single actionable metric for adaptive interference management.

Based on the estimated $\alpha(\theta)$, the BS determines the optimal SIC decoding sequence and instructs the receivers accordingly. Specifically, when $\alpha(\theta)\approx1$ (high elevation angles), satellite decoding is prioritized first to effectively mitigate the strong interference. Conversely, when $\alpha(\theta)\approx0$ (low elevation angles), satellite decoding is either omitted or deferred, thereby reducing unnecessary computational complexity. Complementing this adaptive SIC ordering, we design an RIS codebook that includes terrestrial-priority, satellite-priority, and balanced operation configurations. The appropriate RIS mode is selected at runtime based on the current value of $\alpha(\theta)$, followed by a lightweight gradient-based phase refinement step that efficiently tracks rapid channel variations without the computational burden of full reoptimization.

Simulation results confirm that the proposed adaptive, elevation-aware approach achieves up to a 20\% improvement in the throughput of the terrestrial link compared to static RIS-NOMA schemes. Collectively, the elevation-dependent weighting factor $\alpha(\theta)$, adaptive SIC decoding strategy, and optimized RIS configurations enable robust interference management, resulting in substantial enhancements in the spectral efficiency and connectivity of terrestrial users across a wide range of satellite elevation angles.

The remainder of this paper is organized as follows: Section \ref{sec-Syst-Mod} provides a detailed description of the system model, with a particular focus on the terrestrial downlink and the role of the RIS in the presence of unavoidable satellite interference. Section \ref{sec:Optimizing}  examines the design of the adaptive weighting function. Section \ref{sec:Adaptive-RIS-Reflection} describes how RIS phase configurations are adaptively tuned in real time to manage interference effectively. Finally, the conclusions and key findings of the study are summarized in Section \ref{sec:Conclusion}.

\section{System Model\label{sec-Syst-Mod} }
\begin{figure}
\centering{}\includegraphics[width=3.5in]{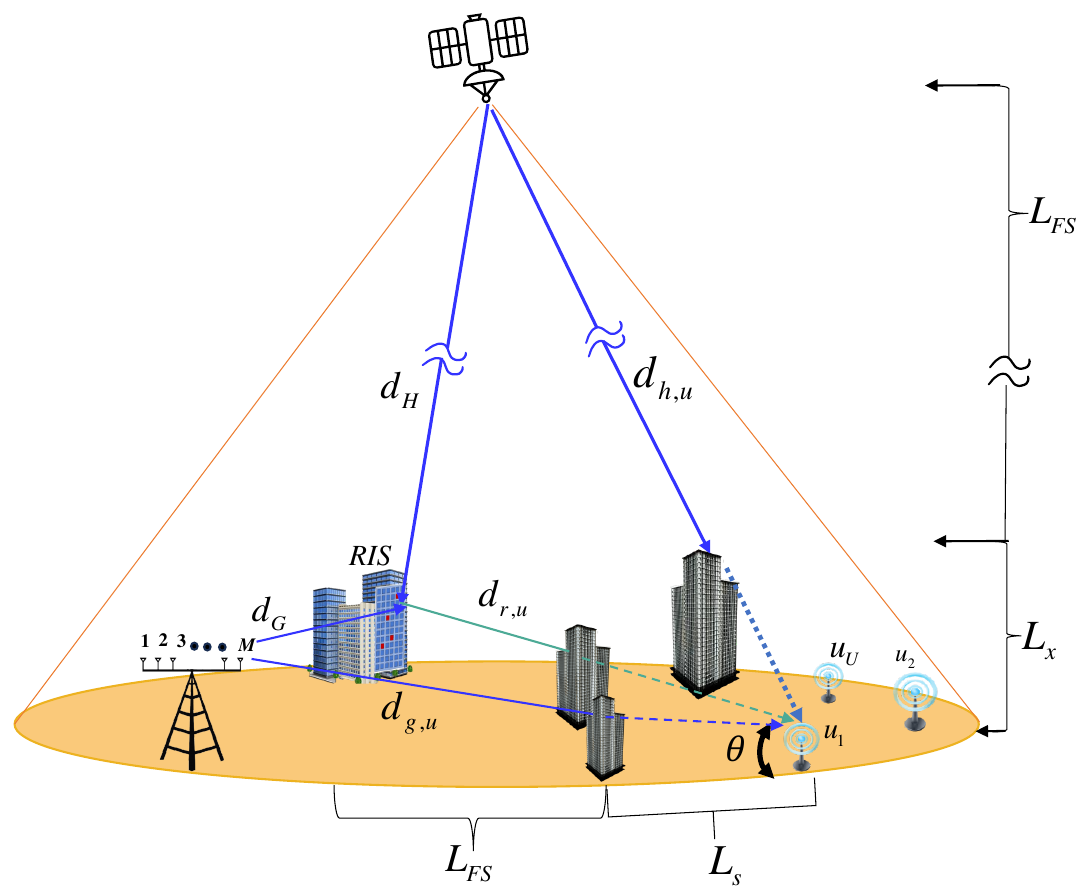}\caption{Architecture of an RIS assisted satellite-terrestrial integrated with
NOMA\label{fig: 1 }}
\end{figure}

We consider the communication scenario illustrated in Fig. \ref{fig: 1 }, where a terrestrial base station (BS) equipped with $M$ antennas serves a group of $U$ single-antenna users, represented by the set $\mathcal{U}=\{1,2,\ldots,U\}$. The BS employs power-domain NOMA, performing superposition coding to transmit signals to multiple users simultaneously. Each user then recovers its own signal using SIC. A RIS is deployed to enhance  signal propagation, while the system also experiences interference originating from satellite transmissions.
The RIS, comprising $N$ passive elements, modulates incoming signals through adjustable phase shifts ($\phi_{n}$). Each element has a reflection coefficient expressed as $\varsigma_{n}=\varGamma_{n}e^{j\phi_{n}}$, with $\varGamma_{n}\thickapprox1$  and $\phi_{n}\in[0,2\pi]$. Collectively, these coefficients constitute a diagonal beamforming matrix $\boldsymbol{\Theta}=\mathrm{diag}([\varsigma_{1},\varsigma_{2} \ldots,\varsigma_{N}])$, which steers reflected terrestrial signals toward users.

Channel models are specified as follows. The terrestrial channel from the BS to the $u$$^{th}$ user is represented
by the direct channel vector $\mathbf{g}_{c,u}\in\mathbb{C}^{M\times1}$, which accounts for urban multipath effects. Similarly, the terrestrial channels between the BS and the RIS elements are described by $\mathbf{G}\in\mathbb{C}^{N\times M}$, and the RIS-to-user channel is represented by $\mathbf{f}_{r,u}\in\mathbb{C}^{N}$, primarily reflecting LoS conditions.

In contrast, the satellite link is modeled to emphasize its role as a source of interfering signals. The channel vector from the satellite to the $u^{th}$ user is denoted by $\mathbf{h}_{u}$, encapsulating shadowed Rician fading, which includes both LoS and diffuse components. Similarly, the satellite-to-RIS channel is represented by $\mathbf{H}\in\mathbb{C}^{N\times1}$. As detailed above, the unsynchronized nature of the satellite downlink with the terrestrial system introduces uncoordinated interference that reduces each NOMA user's signal-to-interference-plus-noise ratio (SINR) and degrades  SIC performance.

The transmitted terrestrial signal from the BS is formulated as
\begin{equation}
\boldsymbol{x_{c}}=\sum_{u=1}^{U}\mathbf{w}_{u}x_{u},
\end{equation}
where $x_{u}$ represents normalized user data   $\left(i.e.,\mathbb{E}\left\{ |x_{u}|^{2}\right\} =1\right) ,$ and beamforming vectors $\mathbf{w}_{u}\in\mathbb{C}^{M\times1}$ are designed under the power constraint $\sum_{i=1}^{U}\|\mathbf{w}_{i}\|^{2}\leq P_{t},$ with $P_{t}$ as the BS's maximum power.

In our model, the satellite downlink signal is treated as an external interference source whose instantaneous power contribution is characterized by a scalar weighting factor $\alpha(\theta)\in\left[0,1\right]$, where $\theta$ denotes the satellite elevation angle. At low elevation angles, the satellite signal traverses a longer, obstructed path, suffering significant path loss and atmospheric attenuation, resulting in an approximately zero value of  $\alpha(\theta)$. As the elevation angle increases, the satellite signal path becomes shorter and predominantly LoS, causing $\alpha(\theta)$  approaching one. Thus, the factor $\alpha(\theta)$ explicitly quantifies the interference strength from the satellite relative to terrestrial signals.

The instantaneous value of $\alpha(\theta)$ is computed by the terrestrial BS based on measured channel gains and the current elevation angle $\theta.$ Using this computed value, the BS determines the optimal SIC decoding order and instructs users accordingly. Specifically, when $0<\alpha\bigl(\theta\bigr)<1$, an adaptive SIC decoding order is selected based on the instantaneous interference strength: if $\alpha(\theta)$ is small, the satellite decoding stage is omitted; if it is moderate, satellite decoding occurs after stronger terrestrial signals have been decoded. Conversely, when $\alpha(\theta)\HF1$, indicating dominant satellite interference, the adaptive rule naturally prioritizes decoding and canceling the satellite interference first.

For benchmarking purposes, we also define a non-adaptive reference scenario in which the BS intentionally assumes a fixed interference condition by setting $\alpha(\theta)=1$ at all elevation angles. This represents a conservative worst-case scenario in which satellite decoding is prioritized first, regardless of the actual elevation-dependent channel conditions.

Formally, the decoding scenarios based on $\alpha(\theta)$ are defined as follows:
\begin{equation}
\alpha\bigl(\theta\bigr)=\begin{cases}
0<\alpha\bigl(\theta\bigr)<1, & \textrm{Adaptive SIC\ensuremath{\left(\textrm{weight-aware}\right)} }\\
\alpha\bigl(\theta\bigr)=1, & \textrm{Non-adaptive SIC ( satellite-first order) }
\end{cases}
\end{equation}
While the BS computes the numerical value of $\alpha(\theta)$  from its channel measurements and satellite elevation angle, our work provides a novel analytical framework that precisely defines how $\alpha(\theta)$ can be modeled, dynamically tuned, and subsequently utilized to drive adaptive interference management and RIS configuration. This framework is essential for transforming complex and time-varying satellite channel conditions into actionable control strategies implemented at the BS.

 Considering these conditions, the received signal at user $u$ is represented by:
\begin{equation}
\begin{gathered}y_{u}\left(\mathbf{\mathit{u},\Theta,}\theta\right)=\alpha\bigl(\theta\bigr)\left(\frac{\mathbf{h}_{u}^{H}}{\sqrt{PL_{s,u}}}+\frac{\mathbf{f}_{r,u}^{H}}{\sqrt{PL_{r,u}}}\mathbf{\Theta}\frac{\mathbf{H}}{\sqrt{PL_{H}}}\right)x_{s}\\
+\sum_{k=1}^{U}\left(\frac{\mathbf{g}_{u}^{H}}{\sqrt{PL_{c,u}}}+\frac{\mathbf{f}_{u}^{H}}{\sqrt{PL_{r,u}}}\mathbf{\Theta}\frac{\mathbf{G}}{\sqrt{PL_{G}}}\right)\mathbf{w}_{k}x_{k}+n_{u},
\end{gathered}
\label{eq: yn}
\end{equation}
where $PL_{s,u},\ensuremath{PL_{r,u}},$ and $PL_{H}$ denote the path losses for satellite-to-user, RIS-to-user, and satellite-to-RIS links, respectively. Similarly, $PL_{c,u}$ and $PL_{G}$ indicate the path losses for the BS-to-user and BS-to-RIS links, respectively. These path-loss values depend on the carrier frequency $f$ and distances $d_{s,u}$, $d_{r,u}$, $d_{H}$, $d_{c,u}$, and $\ensuremath{d_{G}},$ and are formulated based on \cite{3GPP2019} as follows: 
\begin{equation}
PL_{s,u}=L_{FS}(d_{r,u},f)+\underset{\textrm{\ensuremath{\underset{\textrm{\ensuremath{\left(\textrm{between S and }u\right)}}}{\textrm{atmospheric absorption losse }}}}}{\underbrace{L_{\tau}(f,\theta)}}+\underset{\textrm{\ensuremath{\underset{\textrm{\ensuremath{\left(\textrm{user }\textrm{link}\right)}}}{\textrm{shadowing losse }}}}}{\underbrace{L_{x}(\theta)\:,}}
\end{equation}
\begin{equation}
PL_{H}=L_{FS}(d_{H},f)+\underset{\underset{\textrm{\ensuremath{\left(\textrm{between S and RIS}\right)}}}{\textrm{atmospheric absorption losse }}}{\underbrace{L_{\tau}(f,\phi_{n})}}+\underset{\textrm{\ensuremath{\underset{\textrm{\ensuremath{\left(\textrm{RIS }\textrm{link}\right)}}}{\textrm{shadowing losse }}}}}{\underbrace{L_{x}(\phi_{n})},}
\end{equation}
\begin{equation}
PL_{r,u}=L_{\mathit{FS}}(d_{r,u},f)+L_{s}(\theta),
\end{equation}
\begin{equation}
PL_{c,u}=L_{FS}(d_{c,u},f)+L_{s}(\theta),
\end{equation}
\begin{equation}
PL_{G}=L_{FS}(d_{G},f),
\end{equation}
where $L_{\text{FS}}(d,f)=20\log_{10}(f)+20\log_{10}(d)-147.55$ is the free-space path loss, $L_{s}(\theta)$ represents shadowing loss as a function of elevation angle $\theta,$\:$ $$L_{\tau}(f,\cdot)$ denotes atmospheric absorption loss, and $L_{x}(\cdot)$ is an additional loss factor defined in \cite{3GPP2019}. 

The satellite-transmitted signal $x_{s}$ is normalized as $E\left\{ \mid x_{s}\mid^{2}\right\} =1$. The additive white Gaussian
noise at the $u^{th}$ user is modeled as $n_{u}\sim\mathcal{CN}\left(0,\sigma_{u}^{2}\right)$,
where $\sigma_{u}^{2}=\mathtt{\mathfrak{B}\mathtt{\mathcal{K}}T},$ with $\mathfrak{B}$ as the receiver bandwidth, $\mathtt{\mathcal{K}}$ as Boltzmann's constant, and $\mathtt{T}$ as the environmental temperature. 

The SINR at user $u$, which explicitly identifies the contribution of satellite interference, is given by:
\begin{equation}
\text{SINR}(\mathbf{u},\Theta,\theta)=\frac{\left|\mathbb{C}_{u}w_{u}\right|^{2}}{\underset{\text{BS Interference}}{\underbrace{\sum_{\substack{k\neq u}
}\left|\mathbb{C}_{u}w_{k}\right|^{2}}}+\underset{\text{Satellite Interference}}{\underbrace{\alpha^{2}(\theta)\left|\mathbb{S_{\mathit{u}}}\right|^{2}}}+\underset{\text{Noise}}{\underbrace{\sigma_{u}^{2}}}},\label{eq:SNIR}
\end{equation}
where the composite terrestrial channel $\mathbb{C}_{u}$ and the satellite channel $\mathbb{S_{\mathit{u}}}$ are defined as follows:
\begin{gather}
\mathbb{C}_{u}=\frac{\mathbf{g}_{u}^{H}}{\sqrt{PL_{c,u}}}+\frac{\mathbf{f}_{u}^{H}\mathbf{\Theta}\mathbf{G}}{\sqrt{PL_{r,u}PL_{G}}},\nonumber \\
\begin{gathered}\mathbb{S_{\mathit{u}}}=\frac{h_{u}^{H}}{\sqrt{PL_{r,u}}}+\frac{\mathbf{f}_{u}^{H}\mathbf{\Theta}\mathbf{H}}{\sqrt{PL_{r,u}PL_{H}}}.\end{gathered}
\label{eq:}
\end{gather}

In \eqref{eq:SNIR}, the satellite interference term appears as $\alpha^{2}(\theta)\left|\mathbb{S_{\mathit{u}}}\right|^{2}$,
and thus scales quadratically with the elevation-dependent weight $\alpha(\theta)$. Although $\alpha(\theta)$ is determined by the instantaneous channel geometry and environmental attenuation, its value is continuously estimated by the BS via the proposed adaptive  weighting framework. The BS then leverages this real-time $\alpha(\theta)$ to adjust both the SIC decoding order and the RIS configuration, thereby indirectly regulating the satellite's interference contribution to the overall SINR. Such elevation-aware interference management is essential in power-domain NOMA systems for balancing multiple concurrent signals and maximizing spectral efficiency under dynamic channel conditions \cite{Randrianantenaina2020}.

The system capacity for the $u^{th}$ user is derived directly from \eqref{eq:SNIR}, using the Shannon formula:
\begin{equation}
C(u)=B\log_{2}\left(1+\text{SINR}\left(\mathbf{\mathit{u},\Theta,}\theta\right)\right),\label{eq: C_befro}
\end{equation}
where $\mathit{B}$ represents the system bandwidth. 

It is important to note that completely neglecting satellite interference from \eqref{eq:SNIR}, although it may theoretically improve SINR, does not reflect practical scenarios. Satellite links, particularly from LEO constellations, are increasingly integrated into modern networks to provide extended coverage, emergency connectivity, and regulatory compliance \cite{Darwish2022}. Consequently, a realistic system cannot simply disregard satellite signals \cite{Juan2022}. Our proposed adaptive RIS strategy, therefore, offers a practical and effective solution for managing the inevitable satellite interference. 

The integration of satellite links into terrestrial NOMA networks  introduces interference dynamics that are significantly different from purely terrestrial scenarios, complicating conventional SIC. Typically,  terrestrial NOMA systems employ static power-domain superposition, where users decode signals sequentially based on a fixed order determined  by their channel strengths or allocated powers. However, this static ordering becomes inefficient when satellite signals are present due to their unique propagation characteristics, which vary dynamically with elevation angles. Specifically, satellite signals experience elevation-dependent path loss, atmospheric absorption, and shadowing effects, resulting in highly variable interference levels.

The quality of satellite signals, and consequently their interference  contribution, strongly depends on the elevation angle $\theta$. At higher elevation angles ($\theta\to90^{\circ}$), the satellite signal experiences minimal atmospheric attenuation ($L_{\tau}(f,\theta)\thickapprox0$) and negligible shadowing ($L_{x}(\theta)\thickapprox0$), resulting in consistently robust LoS propagation. Conversely, at lower elevation angles, the increased path length, atmospheric absorption ($L_{\tau}(f,\theta)>0$), and significant urban shadowing ($L_{x}(\theta)>0$) severely weaken the satellite signal, making its interference intermittent and less predictable. Such variability challenges conventional static terrestrial SIC approaches, resulting in potential performance degradation and reduced spectral efficiency.

To effectively manage this dynamic interference, we propose an adaptive
SIC ordering approach controlled via the RIS-based weighting factor
$\alpha(\theta)$. Implemented at the RIS, $\alpha(\theta)$ dynamically
scales the satellite's interference contribution in the composite SINR calculation, thereby enabling the adaptive prioritization or suppression of the satellite signal during SIC decoding. Mathematically, the SINR for the user is expressed by \eqref{eq:SNIR}.

At high elevation angles, $\alpha(\theta)$ approaches unity, resulting
in significant satellite interference $\Bigl(\alpha^{2}(\theta)\left|\mathbb{S_{\mathit{u}}}\right|^{2}\Bigr)$. Under these conditions, the receiver prioritizes decoding the satellite signal first within the SIC process, effectively eliminating its interference from subsequent decoding stages. Consequently, terrestrial signals are isolated from satellite-induced distortions, and residual interference is predominantly from terrestrial sources and noise:

\begin{equation}
\mathrm{Residual\:Interference}\propto\sum_{k\neq u}|C_{u}w_{k}|^{2}+\sigma_{u}^{2},
\end{equation}
Conversely, at low elevation angles ($\theta\ll90^{\circ}$), $\alpha(\theta)$ approaches zero, significantly reducing satellite interference $\Bigl(\alpha^{2}(\theta)|S_{u}|^{2}\approx0\Bigr)$. In such scenarios, the receiver adaptively bypasses satellite signal decoding, thereby simplifying the decoding process. This eliminates unnecessary SIC stages associated with decoding weak or unstable satellite signals, reducing complexity, latency, and potential decoding errors. In this regime, the SINR simplifies to a form similar to terrestrial-only NOMA systems:
\begin{equation}
\text{SINR}_{u}\approx\frac{|C_{u}w_{u}|^{2}}{\underset{k\neq u}{\sum}|C_{u}w_{k}|^{2}+\sigma_{u}^{2}}
\end{equation}

The adaptive SIC ordering, governed by the estimated weighting function
$\alpha(\theta)$, ensures that satellite interference is evaluated accurately in real time. Since $\alpha(\theta)$ captures the elevation-dependent interference strength, the BS uses it to select the SIC decoding strategy dynamically. When $\alpha(\theta)\rightarrow1$ at high elevation angles, the satellite component is decoded and canceled first. When $\alpha(\theta)\approx0$ at low angles, satellite decoding is omitted or postponed, thus avoiding unnecessary complexity. This elevation-aware adaptation prevents both overestimation (extra decoding stages) and underestimation (residual interference). The framework aligns with the 3GPP NTN interference management guidelines \cite{Chen2023} and yields two key benefits:
\begin{itemize}
\item At high elevation angles, prioritizing satellite decoding significantly
improves terrestrial SINR by approximately $10\log_{10}\left(1+\frac{\alpha^{2}(\theta)|S_{u}|^{2}}{\sigma^{2}}\right)\text{ dB}$. 
\item At low elevation angles, simplifying the decoding process reduces
latency by eliminating unnecessary satellite decoding stages.
\end{itemize}
 
\section{Design of adaptive elevation-angle-based SIC{\normalsize{} \label{sec:Optimizing}} }

This section describes the design of an adaptive weighting function $\alpha(\theta)$ for managing satellite-induced interference in a RIS-assisted downlink NOMA system.  As detailed in the previous section, the satellite link is modeled exclusively as an unavoidable source of interference rather than as a direct data channel. Because the strength of this interference changes significantly with the satellite's elevation angle, it is crucial to dynamically manage its impact to maintain optimal decoding performance for terrestrial communications. To dynamically manage these variations, the weighting function $\alpha(\theta)$ is implemented at the RIS to effectively scale the satellite interference component during SIC decoding.

\subsection{Adaptive Weighting Function Design}

The adaptive weighting function transitions smoothly from terrestrial dominance at lower elevation angles \textbf{($\theta_{L}$)} to significant satellite interference at higher angles \textbf{($\theta_{H}\rightarrow90^{\circ}$)
}. This transition is represented using a sigmoid function:
\begin{equation}
\alpha\bigl(\theta\bigr)=\frac{1}{1+e^{-r(\theta-\theta_{0})}},\label{eq: Alpa}
\end{equation}
where the parameter $r$ determines the steepness of the transition, and $\theta_{0}$ represents the midpoint elevation angle between terrestrial and satellite-dominated interference conditions. Precisely tuning $r$ ensures a realistic representation of shadowing effects at lower elevation angles and dominant path loss conditions at higher angles, enabling seamless adaptation  as discussed in the following section.

\subsection{Environmental Attenuation Modeling}

Environmental factors impacting signal attenuation are captured by
integrating two statistical models. First, shadowing at low elevation
angles $\theta\:\le\:\theta_{0}$ is modeled as
\begin{equation}
e_{L}(\theta)\sim\mathcal{N}\left(\mu_{L},\sigma_{L}^{2}\right),
\end{equation}
where $\mu_{L}$ and $\sigma_{L}$ represent the mean and standard deviation of shadowing-induced attenuation, respectively. Second, excess path loss at higher elevation angles is modeled using a Gaussian Mixture Model (GMM):
\begin{equation}
e_{\text{G}}(\theta)\sim\sum_{i=1}^{K}\pi_{i}\mathcal{N}(\mu_{\text{H},i},\sigma_{\text{H},i}^{2}),
\end{equation}
where $K$ is the number of Gaussian components, $\pi_{i}$, and $\mu_{\text{H},i}$ and $\sigma_{\text{H},i}$ denote the mixture weight, mean, and standard deviation of each Gaussian component, respectively.

To accurately represent the transition between shadowing-dominated and path-loss-dominated regions, the following combined environmental attenuation model is introduced:
\begin{equation}
\delta(\theta)=\mathit{w}(\theta)\,eL(\theta)+\left[1-w(\theta)\right]\:eG(\theta)
\end{equation}
where the transition weighting function  $\mathit{w}(\theta)$ is defined as:
\begin{equation}
\mathit{w}(\theta)=\frac{1}{1+e^{-\mathcal{K}(\theta-\theta_{0})}}.
\end{equation}
This approach ensures a realistic transition around $\theta_{0}$. Hence, the comprehensive adaptive weighting function that incorporates environmental dynamics is expressed as: 
\begin{equation}
\alpha\bigl(\theta\bigr)=\frac{\left[1+\delta(\theta)\right]}{1+e^{-r(\theta-\theta_{0})}},\label{eq: Alpha_Dlta}
\end{equation}
where $\delta(\theta)$ captures the dynamics of environmental attenuation.

\subsection{Dynamic Tuning Using a PI Controller}
To maintain robust system performance in real time under fluctuating channel conditions, the steepness parameter $r$ is adaptively tuned via a Proportional-Integral (PI) controller:
\begin{equation}
r(t+1)=r(t)+K_{p}\:e(t)+K_{i}\:\tau\:\sum_{\tau}^{t}e(\tau),\label{eq: r_error}
\end{equation}
where the error $e(t)=C_{tar}-C_{obs}(t)$ represents the instantaneous deviation between the target capacity $C_{tar}$ and the observed capacity $C_{obs}(t)$. The proportional gain $K_{p}$ ensures an immediate response to the current error, while the integral gain $K_{i}$ accumulates historical errors, compensating for persistent deviations.

\subsection{Stability Analysis}
The stability of the adaptive mechanism is analyzed using Lyapunov stability theory. A positive definite Lyapunov function $\mathrm{V}(e)=e^{2}$ ensures stability if its time derivative $\dot{\mathrm{V}}(e)=\frac{d}{dt}\mathrm{V}(e)=2e\dot{e}$ remains negative semi-definite. Constraints on $r$ , $K_{p}$ and $K_{i}$ are derived analytically, ensuring:
\begin{equation}
\frac{d\alpha}{d\theta}\Bigg|_{\theta=\theta_{0}}=\frac{\mathtt{\mathit{r}}}{4}\leq\gamma,
\end{equation}
where $\gamma$ is an adaptive learning rate given by:
\begin{equation}
\gamma=\frac{\vartheta}{1+\beta\:|e(t)|},\label{eq:23}
\end{equation}
with  $\vartheta$ and $\beta$ representing the scaling parameters. This ensures stability and prevents excessive corrections that could destabilize the decoding process.

\subsection{Numerical Validation}
To validate the effectiveness of the proposed adaptive $\alpha(\theta)$-based SIC strategy, numerical simulations were carried out using realistic communication parameters aligned with industry standards: carrier frequency $f=2GHz;$ $M=8$ antennas, number of  RIS elements $N=128$, bandwidth $B=10\mathrm{MHz}$ and 4 users. This configuration aligns with 3GPP standards for LEO-based Non-Terrestrial Networks (NTN) \cite{Tuninato2024} and typical Mobile Satellite Service (MSS) channel bandwidths \cite{Kang2024}.

\begin{figure}
\centering{}\includegraphics[width=3.5in]{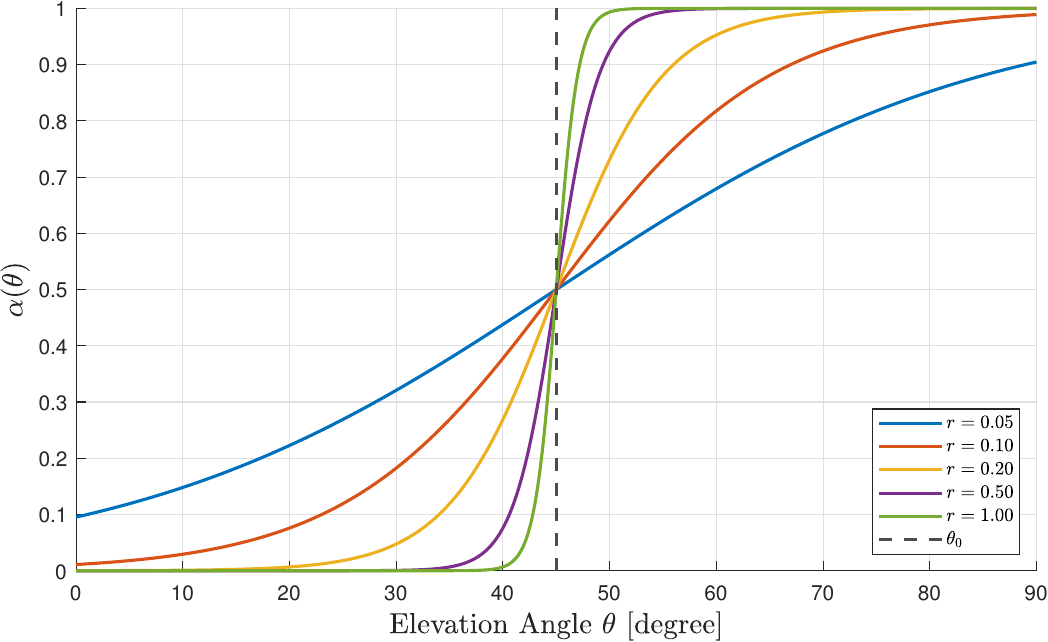}\caption{Effectiveness of dynamic $\alpha(\theta)$ strategy with respect to elevation angles under different $r$ values\foreignlanguage{english}{ \label{fig:2}}}
\end{figure}

\begin{figure}
\centering{}\includegraphics[width=3.5in]{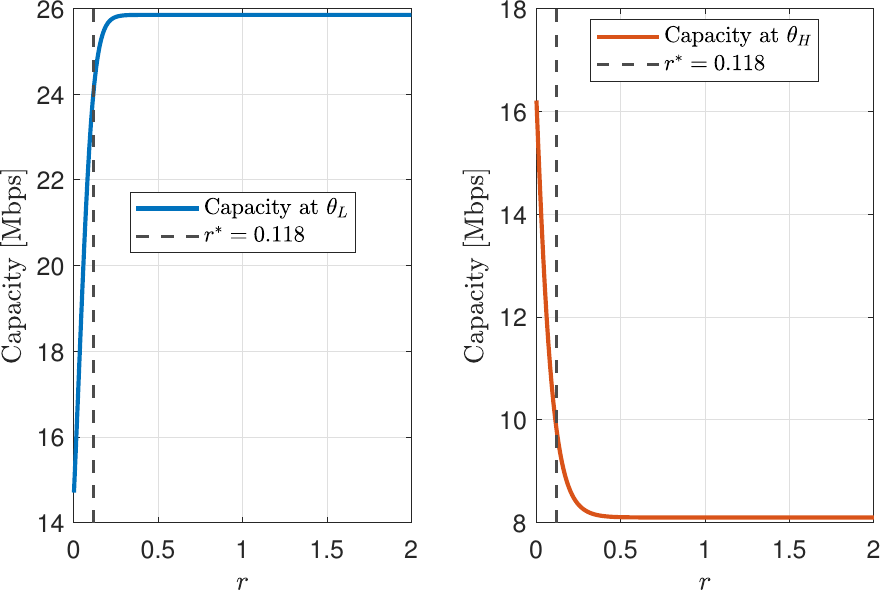}\caption{Impact of steepness parameter $r$ on network capacity at $\theta_{L}$
and $\theta_{H}$, elevation angles\foreignlanguage{english}{ \label{fig: 5 }}}
\end{figure}

Fig. \ref{fig:2} shows how the adaptive weighting function  $\alpha\bigl(\theta\bigr)$ varies with the elevation angle  $\theta$ for different values of the steepness parameter $r$. Smaller $r$ values (such as 0.05 or 0.1) result in a gradual transition of the satellite interference weighting from minimal (at low elevation angles) to substantial (at higher elevation angles). This gradual change is beneficial for smoothly maintaining terrestrial signal dominance while progressively incorporating satellite interference as the elevation angle increases. In contrast, larger $r$ values (e.g. 0.5 or 1.0) produce a rapid transition around the midpoint $\theta_{0}$,
emphasizing satellite interference more abruptly. Thus, these curves highlight the RIS's capability to dynamically and accurately regulate satellite interference in response to varying elevation angles. Among the possible choices, a moderate value of the steepness parameter (e.g., \textbf{$r\approx0.12$}) is found to provide the best tradeoff in practice. This value ensures that satellite interference is incorporated smoothly yet efficiently into the RIS control strategy, effectively preserving terrestrial throughput at low elevation angles while enabling rapid interference suppression as the satellite signal strengthens at higher elevations (see Fig. \ref{fig: 5 } and the associated discussion).

\begin{figure}
\centering{}\includegraphics[width=3.5in]{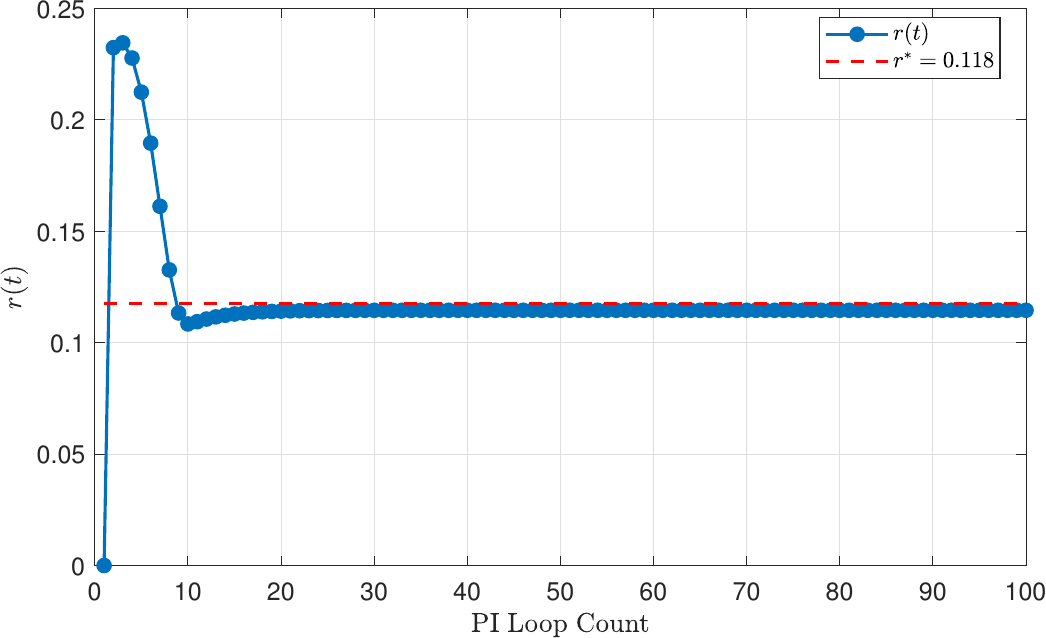}\caption{Dynamic adaptation of the elevation-dependent weighting factor $\alpha$
as a function of elevation angle for different steepness parameter
values $r$ \label{fig:3}}
\end{figure}

To analyze the role of effective interference management within the RIS-assisted NOMA framework, particular attention is directed toward low elevation angles $\theta_{L}$. At these angles, efficient satellite interference suppression is crucial for preserving terrestrial signal integrity during SIC decoding. Although geometric path loss and atmospheric attenuation naturally reduce satellite interference at low angles, residual satellite signals still affect terrestrial signal decoding if not adequately managed. Therefore, we introduce a lower bound \ensuremath{\flat}, on $\alpha(\theta_{L}),$  explicitly constraining the maximum permissible satellite interference level to protect terrestrial communication performance.

Mathematically, the optimal steepness parameter $\overset{*}{r}$ required to enforce this bound at $\theta_{L}$ can be derived by substituting $\theta=\theta_{L}$ into the adaptive weighting expression \eqref{eq: Alpha_Dlta}, yielding: 
\begin{equation}
\overset{*}{r}=-\frac{1}{\theta-\theta_{0}}\ln\left(\frac{1+\delta(\theta)-\flat}{\flat}\right),\label{eq: r_hat}
\end{equation}
which ensures $\alpha(\theta_L)=\flat$.This relationship encapsulates the combined environmental attenuation effects at low elevation angles.

Fig. \ref{fig:3} illustrates the dynamic adaptation of the steepness parameter $r(t)$ over 100 iterations as adjusted by the PI controller. In this example, an upper bound of \ensuremath{\flat}=0.25 is imposed to constrain the satellite contribution at low elevation angles, ensuring that the satellite interference remains controlled during SIC. The PI controller starts with an initial guess of $r(0)=0$ and iteratively updates it to achieve the target system performance. Initially, the parameter $r(t)$ exhibits a transient phase with noticeable oscillations due to the PI correction mechanism. These oscillations gradually dampen, and the system converges to a steady-state value of $r(t)\,\approx\,0.118$, as indicated by the blue line. This converged value matches the theoretical optimum $\overset{*}{r}$ (red dashed line) derived from \eqref{eq: r_hat}. Maintaining $r(t)\,\approx\,\overset{*}{r}$, confirms that the system's real-time adjustments align with the design goal of capping the satellite contribution at 25\% of its full amplitude at low elevation angles. Once stabilized, the PI loop ensures that $\alpha(\theta_L)$ remains close to 0.25, thus satisfying capacity targets without incurring excessive satellite interference. 
\begin{figure}
\centering{}\includegraphics[width=3.5in]{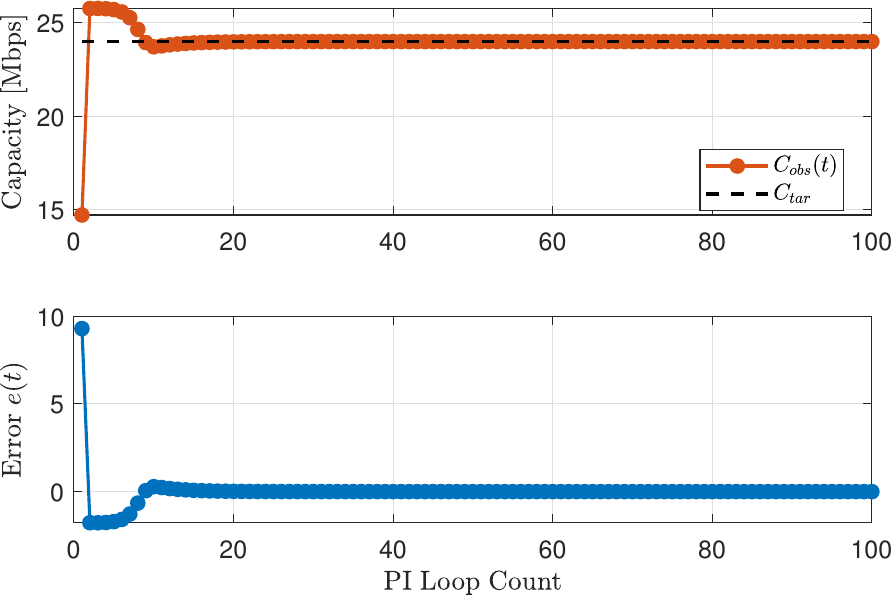}\caption{Dynamic adjustment of the control parameter $r$ in a PI controller
over 100 Iterations (with \ensuremath{\flat}=0.25, \textbf{$\theta_{L}=30^{\circ}$,}
$\theta_{H}=70^{\circ}$ and $C_{tar}$=24 Mpbs) \label{fig:4}}
\end{figure}

Fig. \ref{fig:4} further confirms the adaptive system's effectiveness by displaying the evolution of both the observed system capacity and the corresponding error over 100 PI iterations. A predefined target capacity  $C_{tar}$ = 24  Mbps at the lower elevation angle was set to ensure optimal throughput under constrained satellite interference conditions. Initially, the observed capacity (red curve) overshoots the target due to aggressive controller corrections. However, continuous operation at this higher capacity is undesirable, as it implies that satellite interference is inadequately managed, potentially leading to instability, increased decoding errors, and degraded long-term throughput. After a brief transient period, the PI controller stabilizes the observed capacity around the target of 24 Mbps (dashed line), minimizing the error (bottom subplot). This demonstrates that the adaptive PI controller effectively self-tunes to optimally manage satellite interference, ensuring reliable terrestrial signal decoding at low elevation angles.

Fig. \ref{fig: 5 } provides additional insight into the relationship
between system throughput and the steepness parameter $r$ at two
distinct elevation angles: \textbf{$\theta_{L}=30^{\circ}$ }and \textbf{$\theta_{H}=70^{\circ}$}. At the lower elevation angle (left subplot), capacity increases rapidly from $r=0$, reaching a stable saturation point around the optimal steepness parameter $\overset{*}{r}\approx0.118$. Beyond this optimal point, further increasing $r$ adversely impacts terrestrial performance due to heightened satellite interference, thus providing minimal or no throughput benefit. Conversely, at the higher elevation angle (right subplot), even modest increments in $r$ significantly reduce capacity due to rapidly increasing interference from the satellite link. Thus, these results underscore why the PI controller in Fig. \ref{fig:4} must avoid persistent overshoot: operating consistently above the optimal capacity target can severely degrade long-term terrestrial performance by inadequately mitigating interference. Choosing a steepness parameter around $r\approx0.118$ successfully balances terrestrial throughput at low elevation angles and sustains acceptable performance at higher elevations, representing a robust practical trade-off.

Overall, the numerical results confirm the effectiveness of the proposed adaptive strategy, demonstrating that the RIS-assisted NOMA system can dynamically and precisely manage satellite interference. This enables the system to maintain optimal and stable terrestrial communication performance under a wide range of satellite elevation
conditions.

\section{Adaptive RIS Reconfiguration Based on $\alpha(\theta)$ and Real-Time
Conditions\label{sec:Adaptive-RIS-Reflection}}

Building upon the established satellite and terrestrial weighting function  $\alpha(\theta)$, this section describes how RIS phase configurations \textbf{(\textgreek{J})} can be adaptively tuned in real time to effectively manage interference and optimize signal reception. The core concept involves aligning RIS reflections with the elevation-angle-dependent control function  $\alpha(\theta)$ to dynamically enhance signal quality and reduce interference in a  NOMA-based network as follows.
\begin{itemize}
\item \textbf{\textit{Low elevation angles}}\textit{ }$\left(\alpha(\theta)\approx\:0\right)$:
The RIS emphasizes terrestrial signals from the BS, significantly suppressing satellite interference to ensure clear NOMA decoding without disruption.
\item \textbf{\textit{High elevation angles}} $\left(\alpha(\theta)\approx\:1\right)$:
The satellite link dominates and behaves as a strong and  mostly LoS interferer. In this regime, the RIS is switched to the satellite-priority configuration, tuning its phase shifts to coherently superimpose the direct and reflected satellite signals at each user. This deliberates signal focusing amplifies the satellite component, so that it can be decoded and cancelled first via SIC, thereby minimizing any residual interference on the subsequent terrestrial signal decoding.
\item \textbf{\textit{Intermediate elevation angles}} $\left(0<\alpha(\theta)<1\right)$:
Neither link fully dominates. Hence, RIS operates in a balanced mode, enhancing both terrestrial and satellite signals simultaneously. Iterative or gradient-based optimization methods are utilized to dynamically adjust RIS phase configurations, ensuring optimal signal conditions.
\end{itemize}

To practically implement these modes, a codebook-based method is adopted, consisting of the following three pre-optimized RIS configurations: 
\begin{itemize}
\item $\mathbf{\Theta}_{\text{BS}}$(terrestrial-priority):   Maximizes the aggregated BS\textendash user channel gain; selected when $\alpha(\theta)\approx0$
\item $\mathbf{\Theta}_{\text{Balanced}}$ (mixed-priority): Balances the contributions of terrestrial and satellite links; selected for intermediate
$\alpha(\theta)\in(0,1)$
\item $\mathbf{\Theta}_{\text{SAT}}$(satellite-priority): Aligns RIS phase-shifts to coherently combine direct and reflected satellite signals, enhancing
SIC\textquoteright s ability to cancel the dominant satellite interference when $\alpha(\theta)\approx1$. 
\end{itemize}
These pre-optimized configurations are selected based on real-time $\alpha(\theta)$ evaluations, reducing computational complexity by shifting the intensive optimization process offline.

\subsection{RIS configuration for BS link optimization }

The RIS phase-shift matrix $\mathbf{\Theta}_{\text{BS}}$ is optimized to maximize the received signal power at the user from the BS, ensuring efficient downlink transmission. This configuration addresses the composite channel formed by the direct BS-to-user link and the RIS-reflected BS-to-RIS-to-user link. The optimization problem is formulated as:
\begin{gather}
\mathbf{\Theta}_{\text{BS}}=\arg\max_{\mathbf{\Theta}}\left(\left\Vert \frac{\mathbf{g}_{u}^{H}}{\sqrt{PL_{c,u}}}+\frac{\mathbf{f}_{u}^{H}\mathbf{\Theta}\mathbf{G}}{\sqrt{PL_{r,u}PL_{G}}}\right\Vert ^{2}\right)\label{eq: fi_BS}\\
\text{subject to:}\quad|\Theta_{n,n}|=1,\quad\forall n\nonumber 
\end{gather}
where $\Theta_{n,n}$=$e^{j\phi_{n}}$ represents the phase shifts applied by each RIS element. 

The goal is to maximize the squared norm of the total received signal at the user, which is equivalent to maximizing the power of  received signal. The terms inside the norm represent the superposition of the BS direct path and the BS-RIS path signals. The total signal vector of \eqref{eq: fi_BS} can be expressed using the properties of vector
operations as follows:

\begin{equation}
\left\Vert \mathrm{\boldsymbol{s}}\right\Vert ^{2}=\left\Vert \frac{\mathbf{g}_{u}^{H}}{\sqrt{PL_{c,u}}}\right\Vert ^{2}+\left\Vert \frac{\mathbf{f}_{u}^{H}\mathbf{\Theta}\mathbf{G}}{\sqrt{PL_{r,u}PL_{G}}}\right\Vert ^{2}+2\Re\left\{ \frac{\mathbf{g}_{u}^{H}\left(\frac{\mathbf{f}_{u}^{H}\mathbf{\Theta}\mathbf{G}}{\sqrt{PL_{r,u}PL_{G}}}\right)}{\sqrt{PL_{c,u}}}\right\} ,\label{eq: S}
\end{equation}
where $\Re\{.\}$ denotes the real part. In \eqref{eq: S},  $\mathbf{g}_{u}^{H}$ represents the conjugate transpose of the direct channel from the BS to the user, and $\mathbf{f}_{u}^{H}$ represents the conjugate transpose of the channel from the RIS to the user. The matrix $\mathbf{G}$ is the channel from the BS to the RIS, and $\mathbf{\Theta}$ is a diagonal matrix with diagonal entries $\Theta_{n,n}$. 

To maximize the signal power $\left\Vert \mathrm{\boldsymbol{s}}\right\Vert ^{2}$, particularly the real part of the cross-term in \eqref{eq: S} , which influences constructive or destructive interference, we recall that the magnitude of the inner product satisfies the Cauchy-Schwarz inequality:

\begin{equation}
\left|\mathbf{g}_{u}^{H}\mathbf{f}_{u}^{H}\mathbf{\Theta}\mathbf{G}\right|\leq\left\Vert \mathbf{g}_{u}^{H}\right\Vert .\left\Vert \mathbf{f}_{u}^{H}\mathbf{\Theta}\mathbf{G}\right\Vert 
\end{equation}

The equality in this inequality is achieved when the two vectors are aligned in phase. Therefore, to maximize the inner product and thereby the real part of the cross term, the phase of $\mathbf{g}_{u}^{H}$ must match the phase of $\mathbf{f}_{u}^{H}\mathbf{\Theta}\mathbf{G}$. This alignment requires adjusting $\mathbf{\Theta}$, so that the phase shift of each element $n$ aligns the RIS-mediated signal components with the direct path.

Now, we analyze the specific RIS-mediated contribution from each element
$n$ within the matrix. The matrix is diagonal with elements $\Theta_{n,n}$,
influencing the phase of the signal through each corresponding RIS
element. The element $\mathbf{f}_{u}^{H}(n)\Theta_{n,n}\mathbf{G}(n,:)w_{u}$
reflects the combined effect of the RIS phase shift and the channel
from the BS through the RIS to the user, incorporating a single beamforming
vector $w_{u}.$ Here, $\mathbf{G}(n,:)$ denotes the $n^{th}$ row
of $\mathbf{G}$, representing the channel from all antennas at the
BS to the $n^{th}$ RIS element.

The RIS phase shift $\phi_{n}$ that maximizes constructive interference
is determined by negating the inherent phase of the channel-beamforming
product for each element, ensuring that the phases are aligned to
enhance the signal strength at the user. This is mathematically formulated
as:
\begin{equation}
\phi_{n}=-\angle\left(\mathbf{f}_{u}^{H}(n)\mathbf{G}(n,:)w_{u}\right)
\end{equation}
Consequently, the phase shift for each RIS element $\Theta_{n,n}$ is set to:
\begin{equation}
\Theta_{n,n}=e^{j\phi_{n}}=e^{-\left(\mathbf{f}_{u}^{H}(n)\mathbf{G}(n,:)w_{u}\right)}.\label{eq:f_BS}
\end{equation}
\subsection{RIS configuration for satellite channel optimization ($\mathbf{\Theta}_{\text{SAT}}$)}

At high elevation angles (\textbf{$\theta\thickapprox90^{\circ}$}), the RIS configuration \textbf{$\mathbf{\Theta}_{\text{SAT}}$} focuses specifically on managing satellite interference. Although the satellite link is not utilized as a data transmission channel, its presence as interference is unavoidable. Hence, the RIS phases are optimized to maximize the received satellite interference power at the user. This deliberate maximization may seem unexpected; however, it is essential for effective interference management. 

By maximizing the satellite signal strength at the receiver, the RIS ensures a coherent and constructive combination between the direct satellite signal and the RIS-reflected satellite signal. This enhancement improves the clarity and strength of the satellite signal, making it easier and more accurate for the receiver to identify, decode, and subsequently eliminate this interference through SIC. Precisely removing this interference significantly reduces the residual effect on terrestrial signals, thus enhancing the overall quality of terrestrial signals and the SINR  performance.

Mathematically, thie RIS configuration for satellite interference management is obtained by solving the following optimization problem:
\begin{gather*}
\mathbf{\Theta}_{\text{SAT}}=\arg\max_{\mathbf{\Theta}}\left(\left\Vert \frac{\mathbf{h}_{u}^{H}}{\sqrt{PL_{c,u}}}+\frac{\mathbf{f}_{u}^{H}\mathbf{\Theta}\mathbf{H}}{\sqrt{PL_{r,u}PL_{H}}}\right\Vert ^{2}\right)\\
\text{subject to:}\quad|\Theta_{n,n}|=1,\quad\forall n=1,\dots,N.
\end{gather*}
Here, the objective function represents the total satellite interference arriving at the user through both direct and RIS-reflected paths. Maximizing this quantity enhances the strength and coherence of the received interference signal.

To ensure the coherent superposition of the two interference paths, the optimal RIS phase shifts, $\phi_{n}$, are determined by aligning the reflected interference path with the direct satellite interference path. Consequently, the RIS element phase shifts are calculated as follows:
\begin{equation}
\phi_{n}=-\angle\left(\mathbf{f}_{u}^{H}(n)\mathbf{H}(n)\right),
\end{equation}
 where $\mathbf{H}(n)$ is the channel coefficient from the satellite to the $n^{th}$ RIS element. The corresponding reflection of the RIS element becomes:
\begin{equation}
\Theta_{n,n}=e^{j\phi_{n}}=e^{-\left(\mathbf{f}_{u}^{H}(n)\mathbf{H}(n)\right)},
\end{equation}
By adjusting the RIS elements in this manner, the satellite interference signals are maximized, enabling effective interference identification and cancellation at the receiver. 

\subsection{{\normalsize{}Balanced RIS configuration for joint satellite-terrestrial optimization}} In hybrid satellite-terrestrial networks, especially at intermediate elevation angles, neither the terrestrial nor the satellite interference channel dominates. Therefore, a balanced RIS configuration $\mathbf{\Theta}_{\text{Balanced}}$ is essential to simultaneously enhance the terrestrial data link and optimally manage the unavoidable satellite interference. The RIS operates in this balanced mode when the weighting function $\alpha(\theta)$ lies within an intermediate range specifically designated as $0.3\leq\alpha(\theta)\leq0.7$. The rationale behind selecting these particular threshold values (0.3 and 0.7) is to clearly define operational boundaries: values below 0.3 indicate terrestrial dominance, values above 0.7 signify predominant satellite interference requiring robust cancellation, and intermediate values suggest that neither link is distinctly dominant.

Consequently, for intermediate $\alpha(\theta)$, the RIS configuration is optimized to achieve a balanced contribution, effectively managing interference while ensuring robust terrestrial communication. The optimization problem is defined as:
\begin{align}
\mathbf{\Theta}_{\text{Balanced}} = 
\arg\max_{\mathbf{\Theta}} \bigg( &
\alpha(\theta) \left\Vert \mathbf{h}_{u} + \mathbf{f}_{u}^{H} \mathbf{\Theta} \mathbf{H} \right\Vert^{2} \notag\\
&+ (1 - \alpha(\theta)) \left\Vert \mathbf{g}_{u} + \mathbf{f}_{u}^{H} \mathbf{\Theta} \mathbf{G} \right\Vert^{2} \bigg)
\label{eq:th_balanced} \\
\text{subject to:}\quad & |\Theta_{n,n}| = 1, \quad \forall n = 1,\dots,N. \notag
\end{align}
In this formulation, the weighting function $\alpha(\theta)$ adaptively adjusts the emphasis between the terrestrial link (weighted by $1-\alpha(\theta)$) and satellite interference (weighted by $\alpha(\theta)$), reflecting the relative contributions of these signals based on real-time elevation angle conditions.

The combined contributions from satellite interference and terrestrial signals can be explicitly expressed through the objective function $J(\mathbf{\Theta})$ as:
\begin{flalign}
J(\mathbf{\Theta}) & =\alpha(\theta)\underset{\textrm{Satellite interference contribution}}{\underbrace{\left|\mathbf{h}_{u}+\sum_{n=1}^{N}\mathbf{f}_{u}^{H}(n)e^{j\phi_{n}}\mathbf{H}(n)\right|^{2}}}\nonumber \\
+ & \left(1-\alpha(\theta)\right)\underset{\textrm{Terrestrial data contribution}}{\underbrace{\left|\mathbf{g}_{u}+\sum_{n=1}^{N}\mathbf{f}_{u}^{H}(n)e^{j\phi_{n}}\mathbf{G}(n)\right|^{2}}}\label{eq: J(fi)}
\end{flalign}
where $\mathbf{H}(n)$ and $\mathbf{G}(n)$ represent the channel
coefficients from the satellite and terrestrial BS to the $n^{th}$ RIS element, respectively.

To optimize the objective function $J(\mathbf{\Theta})$, we compute
the gradient with respect to each RIS phase element $\phi_{n}$. Applying the chain rule, the gradient is calculated as:
\begin{gather}
\frac{\partial J}{\partial\phi_{n}}=2\alpha(\theta)\,\Re\left(j\mathbf{f}_{u}^{H}(n)\mathbf{H}(n)e^{j\phi_{n}}\biggl(\mathbf{h}_{u}+\sum_{n=1}^{N}\mathbf{f}_{u}^{H}(n)e^{j\phi_{n}}\mathbf{H}(n)\biggr)^{*}\right)\nonumber \\
+2\left(1-\alpha(\theta)\right)\Re\left(j\mathbf{f}_{u}^{H}(n)\mathbf{G}(n)e^{j\phi_{n}}\biggl(\mathbf{g}_{u}+\sum_{n=1}^{N}\mathbf{f}_{u}^{H}(n)e^{j\phi_{n}}\mathbf{G}(n)\biggr)^{*}\right),\label{eq: jj}
\end{gather}
where $\ensuremath{(\cdot)^{*}}$ denotes complex conjugation. To ensure that each RIS phase shift satisfies the unit modulus constraint $|\Theta_{n,n}|=1$, we project this gradient onto the unit circle, yielding the Riemannian gradient:
\begin{equation}
\mathrm{grad_{\phi_{n}}}=\Re\left(\frac{\partial J}{\partial\phi_{n}}\cdot e^{-j\phi_{n}}\right).
\end{equation}
Each RIS phase $\phi_{n}$ is then updated iteratively through gradient ascent to optimize the balanced RIS configuration:
\begin{equation}
\phi_{n}^{(k+1)}=\phi_{n}^{(k)}+\xi\:\mathrm{grad_{\phi_{n}}},
\end{equation}
where $\xi$ denotes a suitably chosen learning rate that facilitates effective convergence to the optimal solution.

This systematic approach enables the RIS to dynamically adjust its phase configuration, balancing satellite interference management with terrestrial signal enhancement. This ensures robust and reliable decoding in scenarios with intermediate elevation angles. 
%%%%%%%%%%%%%%%%%%%%%%%%%%%%%%%%%%%%%%%%%
\begin{algorithm}[t]
\caption{Balanced RIS Phase Shift Optimization}
\label{alg:balanced_compact}
\KwIn{Channels $\mathbf{H}, \mathbf{G}, \mathbf{f}_u, \mathbf{g}_u$, weight $\alpha$, learning rate $\eta$, iterations $K$.}
\KwOut{Optimized RIS phase shift matrix $\mathbf{\Theta}_{\mathrm{Balanced}}$.}
\BlankLine
Initialize $\phi_n^{(0)} \gets 0$ for all $n = 1,\dots,N$\;
\For{$k \gets 1$ \KwTo $K$}{
  \For{$n \gets 1$ \KwTo $N$}{
    $A_n \gets \mathbf{f}_u^H(n)\,\mathbf{H}(n)$\;
    $B_n \gets \mathbf{f}_u^H(n)\,\mathbf{G}(n,:)\,\mathbf{w}_u$\;
    Compute gradient via Eq.~(33)\;
    Compute Riemannian gradient via Eq.~(34)\;
    Update phase shift via Eq.~(35)\;
  }
}
$\mathbf{\Theta}_{\mathrm{Balanced}} \gets \operatorname{diag}\bigl(e^{j\phi_1^{(K)}},\dots,e^{j\phi_N^{(K)}}\bigr)$\;
\Return{$\mathbf{\Theta}_{\mathrm{Balanced}}$}\;
\end{algorithm}
 %%%%%%%%%%%%%%%%%%%%%%%%%%%%%%%%%%%%%%%%%%%%%%%%%%%%%%%%%
\subsection{\label{subsec:4.4}Adaptive RIS Configuration Selection}

Building on the predefined RIS configurations $\mathbf{\Theta}_{\text{BS}},\mathbf{\Theta}_{\text{Balanced}},\mathbf{\Theta}_{\text{SAT}},$
the system employs a precomputed codebook, optimized offline, to address
the varying interference environment associated with different satellite
elevation angles. These configurations are strategically selected
according to the real-time value of the adaptive weighting function
$\alpha(\theta)$: 

\begin{gather}
\mathbf{\Theta}_{\text{selected}}=\begin{cases}
\mathbf{\Theta}_{\text{BS}}, & \text{if }\alpha(\theta)<0.3,\\
\mathbf{\Theta}_{\text{Balanced}}, & \text{if }0.3\leq\alpha(\theta)<0.7,\\
\mathbf{\Theta}_{\text{SAT}}, & \text{if }\alpha(\theta)\geq0.7.
\end{cases}\label{eq: Th_t}
\end{gather}

This codebook-based selection provides a robust first stage of adaptation
to large-scale channel conditions dictated by elevation angle and
shadowing statistics. However, to address short-term channel variations,
such as those caused by user mobility or fast fading, the initial
configuration $\mathbf{\Theta}_{\text{selected}}$ is further refined
online. This refinement is achieved through a lightweight, real-time  gradient-based update as follows:

\begin{equation}
\mathbf{\Theta}_{\text{final}}=\mathbf{\Theta}_{\text{selected}}+\eta\:\nabla_{\mathbf{\Theta}}\left[\text{SINR}\right]
\end{equation}
where $\eta>0$ is the learning rate, and the update employs the Riemannian
gradient of the SINR with respect to the RIS phases, ensuring the
unit-modulus constraint on all elements is maintained. This two-stage
approach enables the RIS to remain agile in the presence of fast channel
and interference dynamics, while incurring minimal computational overhead. 

In parallel, the SIC decoding order is dynamically adapted according
to $\alpha(\theta)$. When $\alpha(\theta)$ approaches unity (high
elevation), strong satellite interference is anticipated and decoded
first, maximizing the benefit of interference cancellation for terrestrial
users. Conversely, when $\alpha(\theta)$ is near zero (low elevation), the contribution of satellite interference is negligible and is omitted
from the SIC process, thus reducing unnecessary complexity and delay.

The resulting SINR for user $u$ with the refined RIS configuration is expressed as:
\begin{equation}
\text{SINR}(\mathbf{u},\Theta,\theta)=\frac{\left|\acute{\mathbb{C}}_{u}w_{u}\right|^{2}}{\sum_{\substack{k\neq u}
}\left|\acute{\mathbb{C}}_{u}w_{k}\right|^{2}+\alpha^{2}(\theta)\left|\acute{\mathbb{S}}_{u}\right|^{2}+\sigma^{2}}\label{eq:SNIR-1}
\end{equation}
where $\acute{\mathbb{C}}_{u}$ and $\acute{\mathbb{S}}_{u}$ represent
the refined terrestrial and satellite channels, respectively.

By combining codebook-based mode selection with real-time gradient
refinement, the proposed adaptive RIS framework effectively mitigates
both terrestrial and satellite interference across a broad range of
elevation angles and channel conditions. This dual-stage strategy
ensures that the system operates near optimality without incurring excessive complexity.

\begin{figure}
\centering{}\includegraphics[width=3.5in]{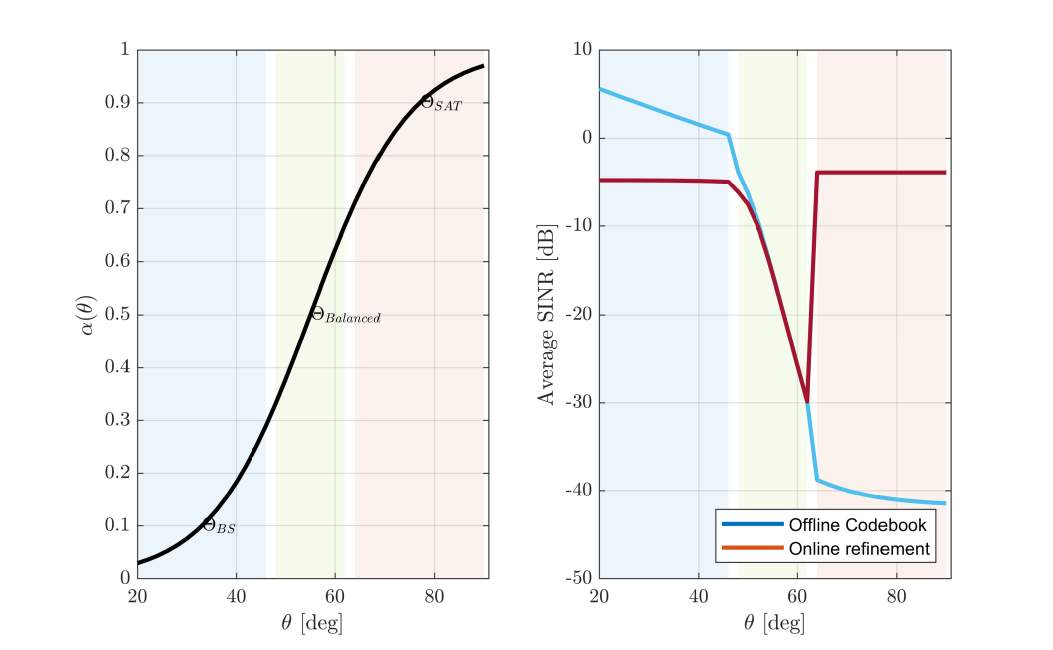}\caption{Elevation-aware RIS configuration: (a) mode boundaries determined
by the elevation-dependent weighting function $\alpha(\theta)$ for
RIS operation. (b) Average SINR versus elevation angle for offline
codebook (blue circles) and online refinement (red squares) approaches,
confirming the necessity of adaptive real-time RIS configuration for
robust satellite-terrestrial integration.\foreignlanguage{english}{\label{fig:6}}}
\end{figure}

Table 1 summarizes the operational regimes and corresponding RIS objectives.
The effectiveness of this approach is illustrated in Fig. \ref{fig:6},
which shows the elevation-dependent SINR performance for both the
offline codebook and adaptive online refinement strategies, using
parameters $N=128,M=8,$ $BW=10MHZ,U=4$, with randomly assigned distances
and a carrier frequency of 2 GHZ. In particular, Fig. \ref{fig:6}
(a) depicts the weighting function $\alpha(\theta)$ as a function
of $\theta$, clearly demarcating the operational boundaries for RIS
mode selection. The shaded regions correspond to terrestrial-priority,
balanced, and satellite-priority regimes, each mapped to a specific
offline codebook configuration. This partitioning is central to the
dual-stage control, enabling the system to select the most suitable
RIS phase setting based on elevation-dependent interference characteristics.

\begin{table}[ht]
\centering
\caption{Operational Modes and Objectives in RIS-Assisted Systems}
\label{table:operational_modes}
\begin{tabularx}{\columnwidth}{|>{\hsize=0.3\hsize}X|>{\hsize=0.2\hsize}X|>{\hsize=0.5\hsize}X|}
\hline
\textbf{Mode} & \textbf{\(\alpha(\theta)\) Range} & \textbf{RIS Objective} \\ \hline
Terrestrial-Priority & \([0, 0.3]\) \newline \(0 < \theta < 45^\circ\) & Maximize BS channel gain \(\|\mathbf{g}_u + \mathbf{f}_u^H \mathbf{\Theta} \mathbf{G}\|^2\) \\ \hline
Balanced & \([0.3, 0.7]\) \newline \(46^\circ \leq \theta < 61^\circ\) & Balance BS and satellite channel gains \\ \hline
Satellite-Priority & \([0.7, 1]\) \newline \(\theta > 61^\circ\) & Maximize satellite channel gain \(\|\mathbf{h}_u + \mathbf{f}_u^H \mathbf{\Theta} \mathbf{H}\|^2\) \\ \hline
\end{tabularx}
\end{table}

Fig. \ref{fig:6} (b) presents the average SINR as a function of elevation
for both the baseline offline codebook method and the proposed online
refinement strategy. At low elevation angles, where terrestrial links
dominate and satellite interference is negligible, the offline codebook
configuration achieves the highest SINR, consistent with its pre-optimized
nature for this regime. The online refinement yields similar performance,
with only marginal differences that reflect local stochastic effects,
as further optimization is not required. As the elevation increases
and the satellite link becomes dominant, the offline codebook approach
fails to adequately suppress the strong satellite interference, leading
to a sharp decline in SINR. In contrast, the adaptive online refinement
exhibits a pronounced SINR recovery at high elevations, reflecting
its ability to dynamically align the RIS and enable effective SIC-based
cancellation of satellite interference. This abrupt improvement confirms
the theoretical predictions from Section \ref{subsec:4.4}, demonstrating
that robust signal quality in integrated satellite-terrestrial NOMA
systems is only achievable through the combination of offline codebook
selection and real-time adaptive RIS refinement.

In alignment with these results, Fig. \ref{fig:7} presents the total
network throughput across the full elevation range, with each regime
mapped to the corresponding RIS control mode according to the value
of the elevation-dependent weighting function $\alpha(\theta)$. Specifically,
the regime where $\alpha(\theta)<0.3$ corresponds to terrestrial-priority
mode, $0.3<\alpha(\theta)<0.7$ to the balanced mode, and $\alpha(\theta)\geq0.7$
to the satellite-priority mode, as defined in Section \ref{subsec:4.4}. 

At low elevation angles ($\theta<50^{\circ}$, terrestrial-priority,
$\alpha(\theta)<0.3$), both the adaptive and static RIS achieve nearly
identical throughput, as satellite interference is negligible and
the RIS configurations are both optimized for the terrestrial channel.
In the intermediate region ($50^{\circ}<\theta<65^{\circ}$, balanced
mode, $0.3\leq\alpha(\theta)<0.7$ ), the adaptive RIS begins to adjust
its phase shifts dynamically to suppress increasing satellite interference,
thus maintaining throughput and outperforming the static RIS, which
is unable to adapt and suffers steadily declining performance. At
high elevation angles ($\theta>65^{\circ}$, satellite-priority, $\alpha(\theta)\geq0.7$),
the adaptive RIS configuration actively aligns to the satellite path,
enabling effective SIC and facilitating a dramatic throughput recovery,
in stark contrast to the static RIS which continues to experience
persistent throughput degradation due to unmitigated interference.

This elevation-dependent behavior is further quantified by the overall
weighted throughput improvement, defined as:

\begin{figure}
\centering{}\includegraphics[width=3.5in]{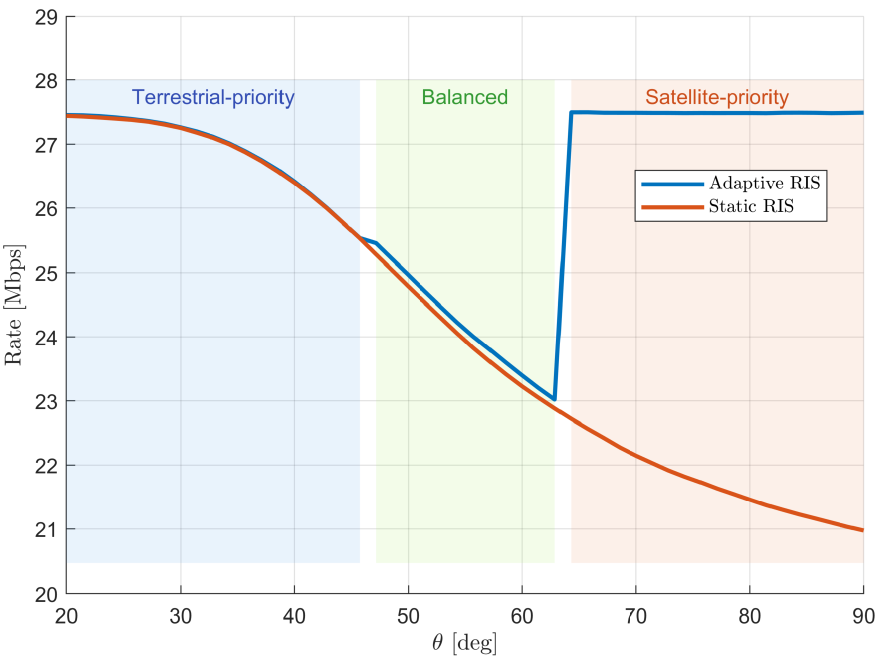}\caption{Total network throughput versus satellite elevation angle with regime-specific
RIS control modes \foreignlanguage{english}{\label{fig:7}}}
\end{figure}

\begin{equation}
\text{ Overall Improvement}=\frac{\sum_{i}\alpha(\theta_{i})\left[R_{\text{static}}(\theta_{i})-R_{\text{adaptive}}(\theta_{i})\right]}{\sum_{i}\alpha(\theta_{i})R_{\text{static}}(\theta_{i})},\label{eq:Ov}
\end{equation}
where $R_{\text{adaptive}}(\theta_{i})$ and $R_{\text{static}}(\theta_{i})$
denote the throughput values at elevation angle $\theta_{i}$ for adaptive and static RIS configurations, respectively. This metric emphasizes performance gains specifically in the satellite-priority regime, where the impact of adaptive control is most pronounced.

The results demonstrate that, in the high elevation (satellite-priority) region, the adaptive RIS-NOMA system achieves up to approximately 20\% greater total throughput compared to the static baseline, highlighting the substantial benefits of combining elevation-aware mode selection with real-time online refinement for robust, interference-resilient satellite-terrestrial communications.

\begin{figure}
\centering{}\includegraphics[width=3.5in]{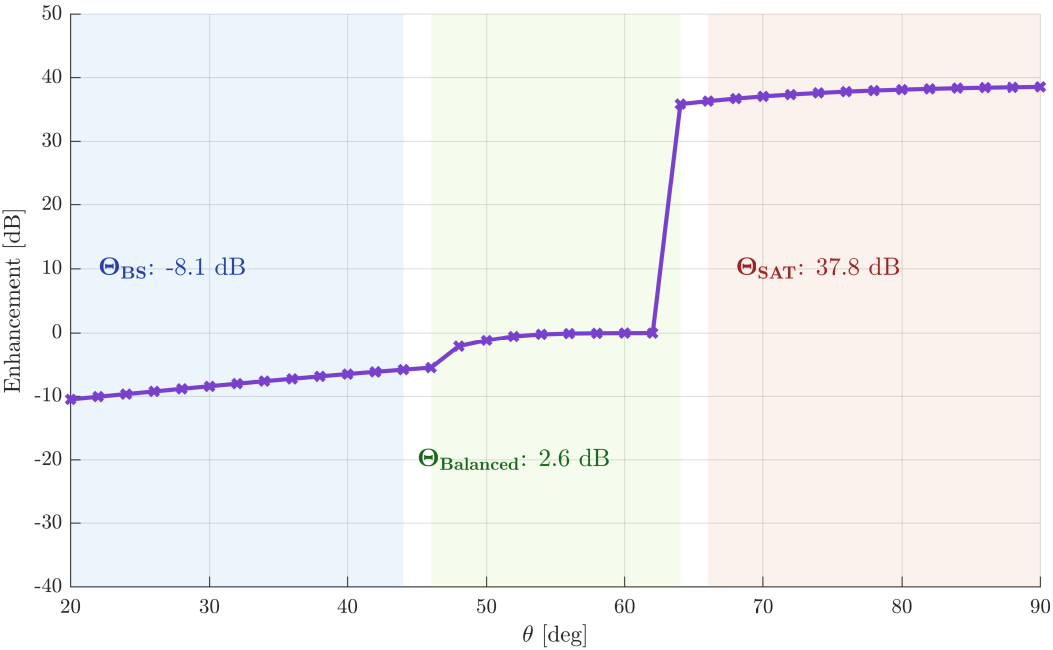}\caption{Elevation-dependent SIC enhancement by adaptive RIS: 7.6 dB (Terrestrial),
2.7 dB (Balanced), and 37.8 dB (satellite) gains over static codebook\label{fig:8}}
\end{figure}

Fig. \ref{fig:8} illustrates the enhancement in SIC performance, measured in dB, achieved by the proposed adaptive RIS configuration with online gradient refinement, relative to a conventional static (codebook-only) RIS scheme as a function of satellite elevation angle $\theta$. This enhancement is computed as the average difference in SINR (in dB) between the adaptive and static RIS cases, following the theoretical model \eqref{eq:SNIR-1}: 
\begin{equation}
\textrm{Enhancemen\ensuremath{t_{u}}[dB]}=10\log_{10}\left(\frac{\sum_{k\ne u}|\acute{\mathbb{C}}_{u}w_{k}|^{2}+|\acute{\mathbb{S}}_{u}|^{2}+\sigma^{2}}{\sum_{k\ne u}|\acute{\mathbb{C}}_{u}w_{k}|^{2}+\alpha^{2}(\theta)|\acute{\mathbb{S}}_{u}|^{2}+\sigma^{2}}\right).
\end{equation}
The plot is divided into three regions: terrestrial-priority, balanced, and satellite-priority, corresponding to the operational RIS codebook modes, with each region annotated by its respective average enhancement.

At lower elevation angles, the plot reveals that the adaptive RIS approach may yield negligible or even slightly negative enhancement compared to the static scheme. This small negative gain arises because satellite interference is already minimal, and the codebook-only RIS configuration is already optimized for the terrestrial channel. Consequently, the additional adaptation offers limited further improvement and may occasionally result in minor fluctuations due to stochastic channel effects. Importantly, this does not degrade terrestrial performance in any meaningful way; rather, it reflects that under benign interference conditions, the simpler static approach suffices, and the adaptive strategy incurs no practical penalty.

As the elevation angle increases, however, the adaptive RIS strategy rapidly outperforms the static scheme by effectively suppressing strong satellite interference, resulting in substantial SINR enhancement. On average, the adaptive RIS approach achieves approximately 11 dB greater SINR enhancement compared to the static configuration, demonstrating its ability to dynamically mitigate both satellite and multi-user interference in integrated satellite-terrestrial networks.

These results confirm the theoretical prediction that real-time RIS adaptation, combined with elevation-aware SIC ordering, provides substantial gains in NOMA decoding performance across all elevation angles.

\section{Conclusion \label{sec:Conclusion} }
In this paper, we present an adaptive interference management framework that effectively integrates RIS-assisted NOMA communications with satellite-terrestrial networks by accurately modeling and managing satellite interference. Central to our proposed approach is an elevation-dependent weighting factor $\alpha(\theta)$, which is analytically modeled using a sigmoid function and dynamically tuned in real time with a PI controller. This weighting factor accurately reflects the actual interference conditions driven by the satellite's elevation angle, capturing severe path loss and shadowing effects through a combination of log-normal shadowing and Gaussian mixture modeling. Leveraging the estimated $\alpha(\theta)$, we further propose an adaptive SIC decoding strategy that prioritizes the decoding of satellite interference when dominant (high elevation angles) and omits unnecessary satellite decoding stages when interference is negligible (low elevation angles). Complementing this SIC adaptation, a codebook-based RIS configuration scheme has been developed, encompassing terrestrial-priority, satellite-priority, and balanced reflection modes, thereby ensuring optimal RIS performance under varying channel conditions.

Simulation results validate the effectiveness of this approach, demonstrating up to a 20\% improvement in overall system throughput compared to conventional static RIS configurations. These findings clearly underscore the potential of adaptive RIS beamforming and elevation-aware NOMA interference management for achieving robust, high-throughput connectivity in next-generation integrated satellite-terrestrial networks. 

\bibliographystyle{IEEEtran}
\addcontentsline{toc}{section}{\refname}\bibliography{RMIT}

\end{document}